\pdfoutput=1 
\documentclass[a4paper,11pt]{article}
\usepackage{pos}
\usepackage{color,graphicx,epsfig}
\usepackage{ifpdf}
\usepackage{amsmath}
\usepackage{bm}
\usepackage{color}
\usepackage[english]{babel}
\usepackage{graphicx}
\usepackage{amsfonts}
\usepackage{braket}
\usepackage{hyperref}
\usepackage{enumerate}

\newcommand{\aso}{\bar\alpha_{s}}
\newcommand{\asb}{\alpha_{s,b}}

\newcommand{\ep}{\epsilon}
\newcommand{\slashed}{\slash \hspace{-0.19cm}}

\newcommand{\be}{\begin{equation}}
\newcommand{\ee}{\end{equation}}
\newcommand{\bea}{\begin{eqnarray}}
\newcommand{\eea}{\end{eqnarray}}

\definecolor{Red}{rgb}{1.,0.,0.}
\definecolor{randomcolour}{rgb}{0.2,0.5,0.7}

\DeclareMathAlphabet\mathbfcal{OMS}{cmsy}{b}{n}

\def\OMIT#1{}

\title{Three-Loop Four-Point Scattering Amplitudes in Massless Gauge Theories}

\author[a]{Amlan Chakraborty}
\author*[b,c]{Giulio Gambuti}

\affiliation[a]{Department of Physics and Astronomy, Michigan State University, East Lansing, Michigan 48824, USA}
\affiliation[b]{Rudolf Peierls Centre for Theoretical Physics, University of Oxford, Clarendon Laboratory, Parks Road, Oxford OX1 3PU}

\affiliation[c]{New College, University of Oxford, Holywell Street, Oxford OX1 3BN, UK}

\emailAdd{chkra69@msu.edu}
\emailAdd{giulio.gambuti@physics.ox.ac.uk}

\abstract{We present novel techniques for the computation of three-loop four-parton scattering amplitudes in full color, non-planar gauge theories.
We elaborate on how the analytic results for these amplitudes can be used to confirm the conjectured infrared structure of QCD up to three loops and how all-orders data can be extracted from their high-energy limit.}

\FullConference{%
  Loops and Legs in Quantum Field Theory - LL2022,\\
  25-30 April, 2022\\
  Ettal, Germany
}

\begin{document}

\maketitle

\section{Introduction}
Scattering amplitudes in QFT serve as a bridge between theory and experiment. QCD computations in particular are essential for the increasingly precise measurements of the parameters of the Standard Model and provide stringent constraints for searches of New Physics. In addition, the computation of new scattering amplitudes adds to our understanding of the more general mathematical structures hiding in QFT.

As the number of external particles and of internal loops grows, so does the complexity of the calculation. Up until a couple years ago, the state of the art for massless four particle scattering was the three-loop four-gluon amplitude in $N=4$ Super Yang Mills (SYM) presented in \cite{Henn:2016jdu}. Extensions to QCD or other less supersymmetric theories, where some of the simplifications due to the (dual) conformal symmetry of $N=4$ are absent, were not possible.

The first analytic results for a QCD three-loop four-point amplitude were obtained for the process $q \bar q \to \gamma \gamma$ and presented in \cite{Caola:2020xup}. Building on this, we review the calculations presented in~\cite{Caola:2021izf,Caola:2022dfa,Caola:2021rqz}, where the computation of all four-parton channels up to three-loops in full QCD was tackled for the first time.
The presence of four colored particles in the external states allows for more complex interactions, involving long range exchanges of color charge between initial and final state. This in turn dictates a richer structure, especially starting a three-loops, where the so-called quadrupole interactions turn on. 

\section{Definitions}
In the following we focus on interactions relevant for QCD, however the techniques described below can be applied to massless gauge theories with different field contents. \\
We consider the processes
\begin{align}\label{eq:process}
g(p_1) + g(p_2)+ g(p_3) + g(p_4)  \rightarrow 0, \notag\\
q(p_1) + \bar q(p_2)+ g(p_3) + g(p_4)  \rightarrow 0, \notag\\
q(p_1) + \bar q(p_2)+ \bar Q(p_3) + Q(p_4)  \rightarrow 0,
\end{align}
where $q$ and $Q$ represent quarks with different flavours.\\
All momenta are incoming and massless 
\begin{equation}\label{eq:mom_cons}
p_1^\mu + p_2^\mu  + p_3^\mu + p_4^\mu = 0, \quad p_i^2 = 0
\end{equation}
and we define the set of Mandelstam invariants
\begin{align}\label{eq:mandelstams}
s = (p_1 + p_2)^2, \quad 
t  = (p_1 + p_3)^2, \quad
u = (p_2 + p_3)^2,
\end{align}
which satisfy the relation $ u \!= \!- t \!-\!s $.
The physical $2 \to 2$ scattering processes correspond to~\eqref{eq:process} after the crossing
$p_{3,4} \! \rightarrow \! -p_{3,4}$ has been performed. In terms of the dimensionless ratio
$x = -t/s$
the physical scattering is associated to the region
\begin{equation} \label{eq:physical_region}
 0<x<1.
\end{equation}
In the following, ultraviolet (UV) and infrared (IR) divergences are regulated in \textit{dimensional regularization}. In particular, we choose to work in the 't Hooft-Veltman scheme (tHV)~\cite{tHooft:1972tcz}.

\section{Decomposition in Color and Lorentz Space} \label{sec:decomp}
Scattering amplitudes involving colored particles in the external states carry color indices and therefore are tensors in color space. The particular space we work in is determined by the external states involved in the interaction. By choosing a basis of this tensor space, one can decompose the amplitudes for the processes~\eqref{eq:process} as follows
\begin{align}\label{eq:physical_amplitude}
\mathcal{A}^{a_1 a_2 a_3 a_4}_{gg\to gg} = 4 \pi \asb \, \sum_{i=1}^6 \mathcal{A}_{4g}^{[i]} \mathcal{C}^{4g}_i\, ,\notag\\
\mathcal{A}^{i_1 i_2 a_3 a_4}_{q\bar q\to gg} = 4 \pi \asb \, \sum_{i=1}^3 \mathcal{A}_{qg}^{[i]} \mathcal{C}^{qg}_i\, ,\notag\\
\mathcal{A}^{i_1 i_2 i_3 i_4}_{q\bar q\to \bar Q Q}= 4 \pi \asb \, \sum_{i=1}^2 \mathcal{A}_{4q}^{[i]} \mathcal{C}^{4q}_i\, ,
\end{align}
where $\asb$ is the bare strong
coupling, $a_n$($i_n$) stands for a $SU(N_c)$ index in the adjoint(fundamental) representation, $\mathcal A^{[i]}$ are
color-ordered \emph{partial amplitudes},
and possible choices for the color basis elements $\{\mathcal C^X_i\}$ reads 
\begin{align}\label{eq:color_structures}
&\mathcal{C}^{4g}_1 = \mathrm{Tr}[T^{a_1} T^{a_2} T^{a_3} T^{a_4}] +  \mathrm{Tr}[T^{a_1} T^{a_4} T^{a_3} T^{a_2}], \quad
\hspace{45pt} \mathcal{C}^{4g}_4 =\mathrm{Tr}[T^{a_1} T^{a_2}]\mathrm{Tr}[T^{a_3} T^{a_4}],\nonumber\\
&\mathcal{C}^{4g}_2 = \mathrm{Tr}[T^{a_1} T^{a_2} T^{a_4} T^{a_3}]+\mathrm{Tr}[T^{a_1} T^{a_3} T^{a_4} T^{a_2}], \quad
\hspace{45pt} \mathcal{C}^{4g}_4 =\mathrm{Tr}[T^{a_1} T^{a_2}]\mathrm{Tr}[T^{a_3} T^{a_4}],\nonumber\\
&\mathcal{C}^{4g}_3 = \mathrm{Tr}[T^{a_1} T^{a_3} T^{a_2} T^{a_4}] +  \mathrm{Tr}[T^{a_1} T^{a_4} T^{a_2} T^{a_3}], \quad
\hspace{45pt} \mathcal{C}^{4g}_6 = \mathrm{Tr}[T^{a_1} T^{a_4}]\mathrm{Tr}[T^{a_2} T^{a_3}], \notag\\[10pt]
&\mathcal{C}^{qg}_{1} = ({{T}^{a_3}}{{T}^{a_4}})_{i_2i_1}, \qquad \qquad
\mathcal{C}^{qg}_{2} = ({{T}^{a_3}}{{T}^{a_4}})_{i_2i_1},  \qquad \qquad
\mathcal{C}^{qg}_{3} = \delta^{a_3a_4}\,\delta_{i_2 i_1} ,\notag\\[10pt]
&\mathcal{C}^{4q}_1 = {\delta}_{ i_1 i_4} {\delta}_{ i_2 i_3}
, \;\;\;\quad\qquad\qquad  \mathcal{C}^{4q}_2 = {\delta}_{ i_1 i_2} {\delta}_{ i_3 i_4}.
\end{align}
Here $T^a_{ij}$ are the fundamental $SU(N_c)$ generators, normalized by $\mathrm{Tr}[T^aT^b]
= \frac{1}{2} \delta^{ab}$. The advantages of this color decomposition are two-fold. The partial amplitudes $ \mathcal{A}^{[i]}$ are (a) independently gauge invariant and (b) not independent under exchanges of external momenta. In fact, one can perform the Feynman-diagrammatic computation only for the partial amplitudes which are independent under crossings of the external momenta and obtain the rest only at the end. For the processes above, this means one can restrict the computation to just $\mathcal{A}_{4g}^{[1]},\mathcal{A}_{4g}^{[4]},\mathcal{A}_{qg}^{[1]},\mathcal{A}_{qg}^{[3]},\mathcal{A}_{4q}^{[1]},\mathcal{A}_{4q}^{[2]}$.
In order to compute the set of independent partial amplitudes, we further decompose them in Lorentz (or spin) space. In particular, invoking the definition of the tHV scheme, which fixes the external states in four space-time dimensions, one can project the partial amplitudes on a set of tensors which are independent of the loop-momenta and are a subset of the ones needed in conventional dimensional regularization (CDR) (see \cite{Peraro:2019cjj,Peraro:2020sfm} for details). In tHV the decomposition reads 
\begin{align}\label{eq:tensor_decomp}
\mathcal{A}_{4g}^{[j]} = \sum_{i=1}^{8} \mathcal{F}^{[j]}_{{4g},i} \: T^{4g}_i, \quad
\mathcal{A}_{qg}^{[j]} = \sum_{i=1}^{4} \mathcal{F}^{[j]}_{{qg},i} \: T^{4g}_i, \quad
\mathcal{A}_{4q}^{[j]} = \sum_{i=1}^{2} \mathcal{F}^{[j]}_{{4q},i} \: T^{4q}_i, 
\end{align}
where the functions $\mathcal{F}^{[j]}_{X,i}$ are called \textit{form factors} and the tensors $T_i^X$ are defined as
\begin{align} \label{eq:Tensors}
&T^{4g}_1 = \epsilon_1 \! \cdot \! p_3\; \epsilon_2 \! \cdot \! p_1\; \epsilon_3 \! \cdot \! p_1\; \epsilon_4 \! \cdot \! p_2 , \nonumber \\
&T^{4g}_2 = \epsilon_1 \! \cdot \! p_3\; \epsilon_2 \! \cdot \! p_1\; \epsilon_3 \! \cdot \! \epsilon_4 , \quad
T^{4g}_3 = \epsilon_1 \! \cdot \! p_3\; \epsilon_3 \! \cdot \! p_1\; \epsilon_2 \! \cdot \! \epsilon_4 , \nonumber \\
&T^{4g}_4 = \epsilon_1 \! \cdot \! p_3\; \epsilon_4 \! \cdot \! p_2\; \epsilon_2 \! \cdot \! \epsilon_3, \quad T_5 = \epsilon_2 \! \cdot \! p_1\; \epsilon_3 \! \cdot \! p_1\; \epsilon_1 \! \cdot \! \epsilon_4 , \nonumber \\
&T^{4g}_6 = \epsilon_2 \! \cdot \! p_1\; \epsilon_4 \! \cdot \! p_2\; \epsilon_1 \! \cdot \! \epsilon_3 , \quad
T^{4g}_7 = \epsilon_3 \! \cdot \! p_1\; \epsilon_4 \! \cdot \! p_2\; \epsilon_1 \! \cdot \! \epsilon_2 , \nonumber \\
&T^{4g}_8 = \epsilon_1 \! \cdot \! \epsilon_2\;  \epsilon_3 \! \cdot \! \epsilon_4+ \epsilon_1 \! \cdot \! \epsilon_4\;  \epsilon_2 \! \cdot \! \epsilon_3 + \epsilon_1 \! \cdot \! \epsilon_3\;  \epsilon_2 \! \cdot \! \epsilon_4 , \notag\\[10pt]
&T^{qg}_1 = \bar{u}(p_2){\slashed \ep}_{3}u(p_1) \,\ep_4\!\cdot\! p_2\,,  \quad \quad \quad\quad\;\;
T^{qg}_2 = \bar{u}(p_2){\slashed \ep}_{4}u(p_1) \,\ep_3\!\cdot\! p_1, \nonumber\\
&T^{qg}_3 = \bar{u}(p_2)\slashed p_{3}u(p_1) \,\ep_3\!\cdot\! p_1\,\ep_4\!\cdot\! p_2, \quad \quad
T^{qg}_4 = \bar{u}(p_2)\slashed p_{3}u(p_1) \,\ep_3\!\cdot\! \ep_4, \notag \\[10pt]
&T^{4q}_1 = {\bar u} (p_2) \gamma_\alpha  u (p_1)  {\bar u} (p_4)  \gamma^\alpha u(p_3) \; ,
\quad
T^{4q}_2 = {\bar u} (p_2)\slashed p_3 u(p_1)   {\bar u} (p_4) \slashed p_2 u(p_3) \;.
\end{align} 
Above, $\epsilon(p_i) = \epsilon_i$ stands for the polarization vector of the  $i$-th external gluon, which satisfies the transversality condition $\epsilon_i \! \cdot \! p_i = 0$.  To obtain the tensor bases above we also gauge fixed the external gluons in the following way: 
\begin{align} \label{eq:gauge_choice}
    & gg \to gg: \quad \quad \ep_1\cdot p_2 = \ep_2\cdot p_3 = \ep_3\cdot p_4 = \ep_4\cdot p_1 = 0, \notag\\
    & q \bar{q} \to gg: \quad \quad  \ep_3\cdot p_2 = \ep_4\cdot p_1 = 0. 
\end{align}
Given the definition of the Lorentz tensors $T^X_i$ above, one can obtain the form factors by introducing a set of $P^X_i$, one for each tensor, so that $\sum_{pol} P^X_i T^X_j= \delta_{ij}$. 
The projectors $P_i^X$ are obtained (see \cite{Peraro:2019cjj,Peraro:2020sfm} for a more thorough description) as 
\begin{equation}
  P^X_i = \sum_j (M^{-1})_{ij}^X {T^X_j}^\dagger,
\end{equation}
where the matrix $M_{ij}$ is defined by 
\begin{equation}
  M_{ij} = \sum_{pol} {T^X_i}^\dagger T^X_j.
\end{equation}
The form factors $\mathcal{F}^{[j]}_{X,i}$, obtained by applying the projectors $P^X_i$ on the amplitudes of the corresponding processes, are Lorentz invariant and carry no color indices. However, in addition to kinematics, they also depend on the size of the gauge group $N_c$ and the number of active quark flavours $n_f$. Schematically, each form factor can be written as 
\begin{equation}
    F^{[j]}_{X,i} = \sum_{n,m} f_{X,i}^{[j],(n,m)}(s,t) \: N_c^n n_f^m, 
\end{equation}
where the functions $f_{X,i}^{[j],(n,m)}$ are independently gauge invariant and only depend on the kinematics. Since any cancellation of gauge-dependent terms must happen within these functions, this extra decomposition of the amplitudes allows one to further parallelize the computation.

\section{Helicity Amplitudes}
In the computation of scattering amplitudes, one wishes to express everything in terms of physical building blocks, which are ultimately expected to be the simplest objects. So far we reduced the computation of gauge theory scattering amplitudes to the evaluation of scalar form (and color) factors. A more physical set of quantities are the helicity amplitudes $\mathcal{A}_{\bm \lambda}$, $i.e.$ scattering amplitudes where the helicities of the external states are fixed to ${\bm\lambda} = \{\lambda_1, \lambda_2, \lambda_3, \lambda_4\}$. Because we work in tHV, the number of parity-independent helicity amplitudes for each process in~\eqref{eq:process} is equal to the dimension of the corresponding tensor basis in~\eqref{eq:Tensors}. Indeed, the parity-independent helicity configurations for the processes~\eqref{eq:process} are
\begin{align}
    & gg \to gg: \{++++\},\{-+++\},\{+-++\},\{++-+\},\{+++-\},\{++--\},\{+-+-\},\{+--+\}, \notag\\
    & q \bar q \to gg: \{-+--\},\{-+-+\},\{-++-\},\{-+++\}, \notag\\
    & q \bar q \to \bar Q Q: \{+-+-\},\{+--+\}
\end{align}
and any helicity amplitude corresponding to one of these configurations can be written as a linear combination of the form factors defined above. To make this more explicit, we use the  \textit{spinor-helicity formalism}~\cite{Dixon:1996wi} to fix the helicities of the external gluons and quarks. 
In our conventions, the fixed-helicity gluon and quark polarization vectors take the form 
\begin{equation}\label{eq:polvec}
\epsilon_{i,+}^\mu = \frac{[q_i|\gamma^\mu|i\rangle}{\sqrt{2} [i|q_i]}, \quad
\epsilon_{i,-}^\mu = \frac{[i|\gamma^\mu|q_i\rangle}{\sqrt{2} \langle q_i|i\rangle}, \quad
u_i^+ = |i\rangle, \quad
u_i^- = |i], \quad
\bar u_i^+ = \langle i|, \quad
\bar u_i^- = [i|,
\end{equation}
where the gluon reference momenta $q_i$ are implied by the gauge choices specified in~\eqref{eq:gauge_choice}.
Plugging eq.~\eqref{eq:polvec} in eq.~\eqref{eq:tensor_decomp} and stripping out an overall helicity-dependent spinor factor $s_{\bm \lambda}$, we can write the color-ordered partial amplitudes as
\begin{equation}\label{eq:spinor_factorisation}
\mathcal{A}^{[i]}_{\bm{\lambda}}  = \mathcal{H}^{[i]}_{\bm{\lambda}}   \; s_{\bm{\lambda} },
\end{equation}
where we omitted the process dependence for clarity. The spinor-stripped helicity amplitudes $\mathcal{H}^{[i]}_{\bm{\lambda}}$ can then be written as linear combinations of the form factors $\mathcal{F}^{[j]}_{i}$, where the coefficients are rational functions of $s$, $t$ and $d$. \\

Having identified the $\mathcal{H}^{[i]}_{\bm{\lambda}}$ as "physical" quantities, we can now compute them perturbatively. Details on the computation of the $gg \to gg, q \bar q \to gg$ and $q \bar q \to \bar Q Q$ channels up to three loops can be found on \cite{Caola:2021izf,Caola:2022dfa,Caola:2021rqz}. To sum up, there are two major bottlenecks in these computations: (1) performing the Dirac and color algebra arising from Feynman diagrams; (2) reducing the large number of Feynman integrals ($O(10^7)$ for $gg \to gg$) to a basis of master integrals and subsequently including of the integration-by-parts (IBPs) identities in the amplitude. The amplitude decompositions discussed in Section \ref{sec:decomp} allow one to perform the steps above on smaller independent building blocks, making the computation more manageable. \\

Analytic expressions for the master integrals involved up to three loops are available \cite{Henn:2020lye,Bargiela:2021wuy} as a Laurent series in $\ep$ up to $\mathcal O(\ep^0)$ in terms of HPLs up to transcendental weight six and can be inserted in the expressions for the \emph{bare} helicity amplitudes obtained from Feynman diagrams.

\section{The IR structure}
\begin{figure}
 \centering
    \includegraphics[width=0.25\linewidth]{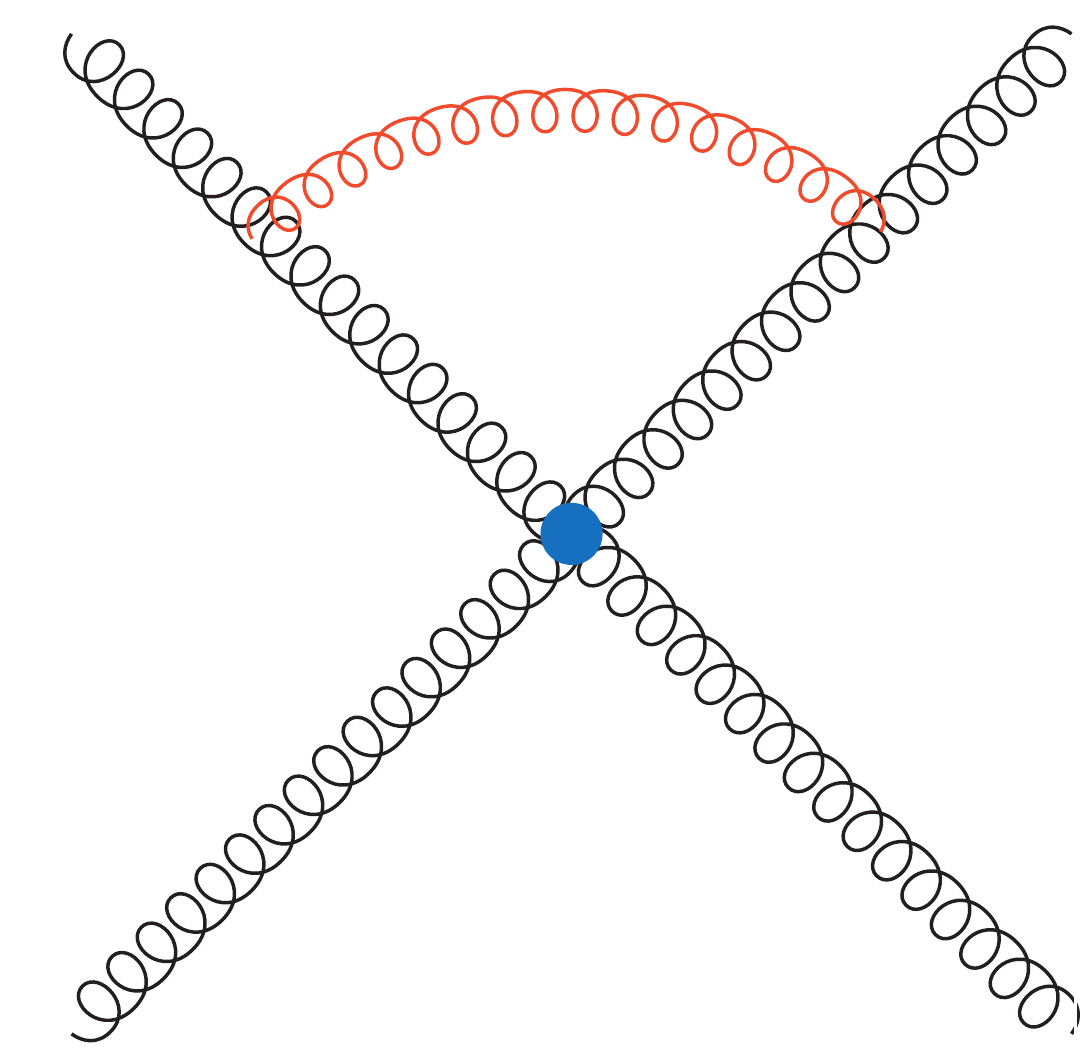}
    \caption{Sample diagram contributing to the dipole matrix $\mathbf{\Gamma}_{\text{dip}}$ for the process $gg \to gg$. The gluon in red is to be thought of as a soft gluon being exchanged by the hard legs (in black).}  \label{fig:dipole}
\end{figure}

After renormalization, the helicity amplitudes 
contain only IR divergencies, which appear as poles in the series expansions of the dimensional regulator $\ep$, and depend on the renormlization scale $\mu$.
Up to three loops, one can factorize the IR physics from the hard scattering and write~\cite{Becher:2009cu,Becher:2009qa,Sterman:2002qn,Aybat:2006wq,Aybat:2006mz,Dixon:2009gx,Gardi:2009qi,Gardi:2009zv,Almelid:2015jia}
\begin{equation}\label{eq:IR_factorisation}
\mathcal{H}_{{\bm{\lambda}},\: \text{ren}} = \mathbfcal{Z}_{IR} \; \mathcal{H}_{{\bm{\lambda}},\: \text{fin}} \; ,
\end{equation}
where  $\mathbfcal Z_{IR}$ is a color matrix that acts on the color
basis $\{\mathcal C_i^X\}$~\eqref{eq:color_structures} and can be
written in exponential form in terms of the soft anomalous dimension $\mathbf\Gamma = \mathbf{\Gamma}_{\text{dip}}  + \mathbf{\Delta}_4  $ as
\begin{equation}\label{eq:exponentiation}
\mathbfcal{Z}_{IR} = \exp \left[ \int_\mu^\infty \frac{\mathrm{d} \mu'}{\mu'}  \mathbf{\Gamma}(\{p\},\mu')\right] .
\end{equation}
Above, the \emph{dipole} term $\mathbf{\Gamma}_{\text{dip}}$ is associated to a long-range pairwise exchange of color charge between external legs (see Figure \ref{fig:dipole}) and explicitly reads
\begin{align}\label{eq:dipole}
\mathbf{\Gamma}_{\text{dip}}  &=  \sum_{1\leq i < j \leq 4} \mathbf{T}^a_i \; \mathbf{T}^a_j\; \gamma^\text{K} \; \ln{\left(\frac{\mu^2}{-s_{ij}-i \delta}\right)}  + \sum_i \gamma^i \; ,
\end{align}
where $s_{ij} = 2p_i \!\cdot\! p_j$, $\gamma^{\text{K}}$ is the \textit{cusp anomalous dimension}
\cite{Korchemsky:1987wg,Moch:2004pa,Vogt:2004mw,Grozin:2014hna,Henn:2019swt,Huber:2019fxe,vonManteuffel:2020vjv} and $\gamma^{i=g,q}$ is the \textit{gluon(quark) anomalous dimension} \cite{Ravindran:2004mb,Moch:2005id,Moch:2005tm,Agarwal:2021zft}.

\begin{figure}[thp]
\centering
    \includegraphics[width=1\linewidth]{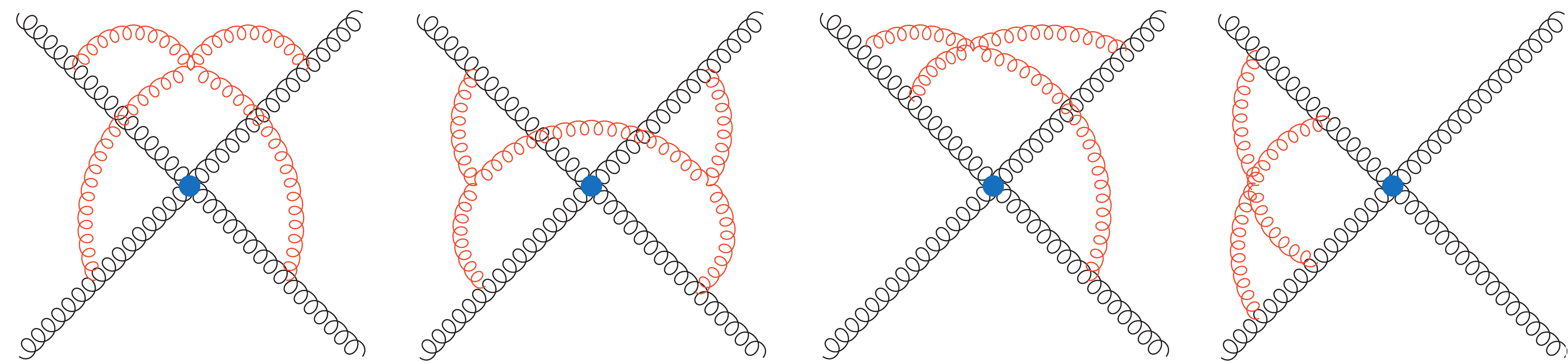}
    \caption{Sample diagrams contributing to the quadrupole matrix $\mathbf{\Delta}^{(3)}_4$ for the process $gg \to gg$. Gluons in red are to be thought of as soft gluons being exchanged by the hard legs (in black).}  \label{fig:quadrupole}
\end{figure}

The \emph{quadrupole} matrix ${\bm \Delta}_4$ is instead due to exchanges of color charge among (up to) four external legs (see Figure \ref{fig:quadrupole}) and it starts playing a role at three loops, where we only need the first order in its perturbative expansion $ \mathbf{\Delta}_4
= \sum_{n=3}^\infty \aso^n \mathbf{\Delta}^{(n)}_4 $. This reads~\cite{Almelid:2015jia}
\begin{align} \label{eq:quadrupole}
&\mathbf{\Delta}^{(3)}_4 = f_{abe} f_{cde}\bigg[
- 16 \,C   \, \sum_{i=1}^4 \sum_{\substack{1\leq j < k \leq4 \\ j,k\neq i}}  \left\{ \mathbf{T}^a_i,\mathbf{T}^d_i \right\} \mathbf{T}^b_j \mathbf{T}^c_k 
\nonumber \\& + 
128 \left[ \mathbf{T}^a_1\mathbf{T}^c_2\mathbf{T}^b_3\mathbf{T}^d_4D_1(x) - \mathbf{T}^a_4\mathbf{T}^b_1\mathbf{T}^c_2\mathbf{T}^d_3 D_2(x) \right]
\bigg],
\end{align}
with  $C = \zeta_5 + 2 \zeta_2 \zeta_3$ and the functions $D_1(x)$ and $D_2(x)$ can be found in~\cite{Caola:2021izf,Caola:2022dfa,Caola:2021rqz}.\\

In the formulas above, $\mathbf{T}^a_i$ represents the color generator of the $i$-th parton in the scattering amplitude:
\begin{alignat}{2}\label{convention}
(\mathbf{T}^a_i)_{\alpha\beta} &=  t^a_{\alpha \beta} \; &&\text{ for a final(initial)-state quark (anti-quark)}, \nonumber\\
(\mathbf{T}^a_i)_{\alpha\beta} &=  -t^a_{\beta\alpha} \; &&\text{ for a final(initial)-state anti-quark (quark)}, \nonumber\\
(\mathbf{T}^a_i)_{bc} &=  -if^{abc} \;
&&\text{ for a gluon.}
\end{alignat}

The computations presented here and in~\cite{Caola:2021izf,Caola:2022dfa,Caola:2021rqz} allowed us to verify up to three loops the structure of IR singularities in QCD predicted by the equations above. This provides a highly non-trivial check of our results and of the techniques described here. 

\section{High Energy Limit and the Gluon Regge trajectory}
Analytic results for the processes~\eqref{eq:process} up to three loops allow us to directly compute the high-energy limit of their amplitudes and peak into the all-orders structure of QCD. Employing the analytic continuation procedure described in \cite{Caola:2021rqz}, we can obtain the scattering amplitudes for the processes
\begin{align}\label{eq:process_regge}
g(p_1) + g(p_2) \rightarrow  g(p_3) + g(p_4) , \notag\\
q(p_1) + g(p_2) \rightarrow q(p_3) + g(p_4)  , \notag\\
q(p_1) + Q(p_2) \rightarrow \bar q(p_3) + Q(p_4),
\end{align}
from the ones described in the previous Sections.
At tree level, the processes in~\eqref{eq:process_regge} are mediated by a gluon exchange and exhibit interesting factorization properties in the limit  $|s| \approx |u| \gg |t|  $, or equivalently $x \rightarrow 0 $. This is the so called \textit{Regge limit}.
To describe the physics in this kinematical region, it is convenient to introduce amplitudes which have definite signature under $s \leftrightarrow u$ exchange:
\begin{equation}
  \mathcal{H}_{\mathrm{ren,\pm}} = \frac{1}{2}\left.[ \mathcal{H}_{\mathrm{ren}}(s,u)   \pm   \mathcal{H}_{\mathrm{ren}}(u,s)  \right]  \,.
\end{equation}
It is also practical to write the logarithmic components of the Regge-limit amplitudes in terms of the signature-even combination of logarithms
\begin{equation}
L = -\ln(x) -  \frac{i\pi}{2}\approx \frac{1}{2}  \left( \ln\left( { \frac{-s-i \delta }{-t} } \right) + \ln\left( {\frac{-u-i \delta }{-t} } \right)  \right).  
\end{equation}
At leading power in $x$ and next-to-leading logarithmic 
(NLL) accuracy (in powers of $L$), the odd
amplitude has a nicely factorized form: to all orders in the strong coupling, $\mathcal H_{\rm ren,-}$ can
be thought of as the amplitude for the exchange of a single ``reggeized'' $t$-channel gluon, whose interactions with the external
high-energy partons are encoded by so-called \emph{impact factors}~\cite{Lipatov:1976zz,Kuraev:1976ge,Fadin:1993wh,Collins:1977jy,Gribov:2009zz}.
This single-particle exchange is referred to as the ``Regge-pole'' contribution to the high-energy limit amplitude and is described by the gluon Regge trajectory $\tau_g = \sum_{n=1} \aso^n \tau_n $.

Beginning at next-to-next-to-leading logarithmic (NNLL) accuracy,  the factorisation of the odd amplitude is broken by the appearance of multiple Reggeon exchanges between the external projectiles~\cite{Gribov:2009zz,DelDuca:2001gu,DelDuca:2008pj,Caron-Huot:2013fea,Caron-Huot:2017fxr,Fadin:2016wso,Fadin:2017nka,Falcioni:2021buo}, which are usually referred to as the ``Regge-cut'' contributions. With the appearance of Regge-cuts alongside the Regge-pole, it becomes non-trivial to disentagle the two contributions. This issue was described in~\cite{Falcioni:2021buo}, where a scheme for separating the two was also presented. 

The three-loop calculations presented here and in~\cite{Caola:2021izf,Caola:2022dfa,Caola:2021rqz} allowed us to compute the last missing ingredient
for the characterization of the signature
even/odd amplitudes at NLL/NNLL, $i.e.$ the three loop Regge trajectory, and check Regge factorisation to this accuracy across all partonic channels for $2 \to 2$ scattering in QCD.

\section{Conclusion}
We described the techinques used for the first computation of helicity amplitudes for the scattering of four partons up to three loops in full QCD.  The methods described are applicable to $2 \to 2$ scattering processes in any massless gauge theory. We reviewed the IR factorization and its practical application to the regularization of these scattering amplitudes, with emphasis on the color quadrupole radiation contribution to the IR anomalous dimension matrix, which we were able to check by direct calculation. We also touched on the Regge limit of these amplitudes, from where we extracted the three-loop Regge trajectory, unlocking the NNLL description of single-Reggeon exchanges in QCD.\\

\noindent {\bf Acknowledgments} \; 
The Authors are thankful to Fabrizio Caola, Andreas von Manteuffel and Lorenzo Tancredi for collaborating on the papers summarized in these proceedings.

GG's research and attendance to the conference ``Loops and Legs in Quantum Field Theory 2022'' was supported by the Royal Society grant URF/R1/191125.

\bibliographystyle{bibliostyle}
\bibliography{biblio}

\end{document}